\documentclass[
superscriptaddress,
nofootinbib,
twocolumn,
 amsmath,amssymb,
 aps,
pra,
notitlepage,
longbibliography
]{revtex4-1}

\usepackage{dcolumn}% Align table columns on decimal point
\usepackage{bm}% bold math
\usepackage{epsfig}
\usepackage{graphicx}
\usepackage{latexsym}
\usepackage{amsfonts}
\usepackage{setspace}
\usepackage{graphicx}
\usepackage{amsmath}
\usepackage{mathrsfs}
\usepackage{verbatim}
\usepackage{color}
\usepackage{SIunits}
\usepackage{hyperref}
\usepackage{tikz}
\usepackage[compat=1.1.0]{tikz-feynman}

\usepackage{physics}
\newcommand{\be}{\begin{equation}}
\newcommand{\ee}{\end{equation}}
\newcommand{\ba}{\begin{array}}
\newcommand{\ea}{\end{array}}
\newcommand{\bqa}{\begin{eqnarray}}
\newcommand{\eqa}{\end{eqnarray}}

\newcommand{\mr}{\mathrm}

\begin{document}

\title{Observation of photon-phonon correlations via dissipative filtering}

\author{Mengdi Zhao} 
\affiliation{Holonyak Micro and Nanotechnology Laboratory, University of Illinois at Urbana-Champaign, Urbana, IL 61801 USA}
\affiliation{Department of Physics, University of Illinois at Urbana-Champaign, Urbana, IL 61801 USA}
\affiliation{Illinois Quantum Information Science and Technology Center (IQUIST), University of Illinois at Urbana-Champaign, Urbana, IL 61801 USA}
\author{Kejie Fang} 
\email{kfang3@illinois.edu}
%\homepage{https://fang.ece.illinois.edu}
%\thanks{These authors contributed equally to this work.}
\affiliation{Holonyak Micro and Nanotechnology Laboratory, University of Illinois at Urbana-Champaign, Urbana, IL 61801 USA}
\affiliation{Illinois Quantum Information Science and Technology Center (IQUIST), University of Illinois at Urbana-Champaign, Urbana, IL 61801 USA}
\affiliation{Department of Electrical and Computer Engineering, University of Illinois at Urbana-Champaign, Urbana, IL 61801 USA}

\begin{abstract} 

Cavity-optomechanics enables photon-phonon interaction and correlations by harnessing the radiation-pressure force. Here, we realize a ``cavity-in-a-membrane'' optomechanical architecture which allows detection of the motion of lithographically-defined, ultrathin membranes via an integrated optical cavity. Using a dissipative filtering method, we are able to eliminate the probe light \emph{in situ} and observe photon-phonon correlations associated with the low-frequency membrane mechanical mode. The developed method is generally applicable for study of low-frequency light scattering processes where conventional frequency-selective filtering is unfeasible. 

\end{abstract}
%\pacs{}

\maketitle

\section{Introduction}

Optomechanical cavities, involving radiation-pressure-coupled electromagnetic fields and mechanical oscillators, enable measurement and control of acoustic excitations, i.e., phonons, using optical means \cite{aspelmeyer2014cavity}. Recently, with the phonon-counting technique \cite{cohen2015phonon,meenehan2015pulsed},  quantum correlations between photons and phonons in optomechanical cavities have been revealed \cite{riedinger2016non}, leading to demonstrations of a mechanical quantum memory \cite{wallucks2020quantum} and optomechanical teleportation \cite{fiaschi2021optomechanical}. The phonon-counting technique essentially uses a pump-probe approach for coherent conversion between phonons and photons followed by frequency filtering to selectively address the acoustically-scattered photons. Due to the need of mechanical sideband filtering, this method is most effective for optomechanical systems with high-frequency mechanical modes---typically in the microwave band. On the other hand, a growing number of optomechanical systems involving mechanical resonators with ultralow losses have emerged for applications ranging from sensing to quantum transduction \cite{reinhardt2016ultralow,norte2016mechanical,tsaturyan2017ultracoherent,ghadimi2018elastic,guo2019feedback}. These low-loss mechanical resonators, relying on specific dissipation and stress engineering, have a frequency typically in the megahertz range and below. As a result, the phonon counting technique involving frequency-selective optical sideband filtering becomes difficult for these low-frequency optomechanical systems \cite{galinskiy2020phonon}.

Here, we develop a tunable dissipative filtering method, which enables attenuation of the transmission of probe photons while simultaneously optimizes the emission of mechanically-scattered photons from a waveguide-coupled optomechanical cavity with low-frequency mechanical modes. In contrast to the frequency-selective filtering, our method based on cavity dissipation engineering separates external probe photons from intra-cavity scattered photons regardless of their frequency. 
We apply the dissipative filtering method to an optomechanical system with a waveguide-coupled ring resonator embedded in a free-standing membrane and observe photon-phonon correlations by photon counting. Such ``cavity-in-a-membrane'' architecture enables detection of membrane vibrational modes with an integrated optical cavity by concentrating optical energy within the wavelength distance of the membrane, in contrast to previous methods utilizing free-space Fabry-Perot cavities \cite{thompson2008strong, higginbotham2018harnessing}. Because the membrane is defined during the lithographic fabrication process, ultrathin membranes with a thickness down to several atomic layers can be made, which might enable new optomechanical sensing and transduction modalities.  The dissipative filtering method developed here is generally applicable to other types of optomechanical systems as well as study of low-frequency light scattering processes, such as dynamic light scattering by macromolecules \cite{stetefeld2016dynamic}, where conventional frequency-selective filtering is unfeasible.

\begin{figure*}[!htb]
\begin{center}
\includegraphics[width=2\columnwidth]{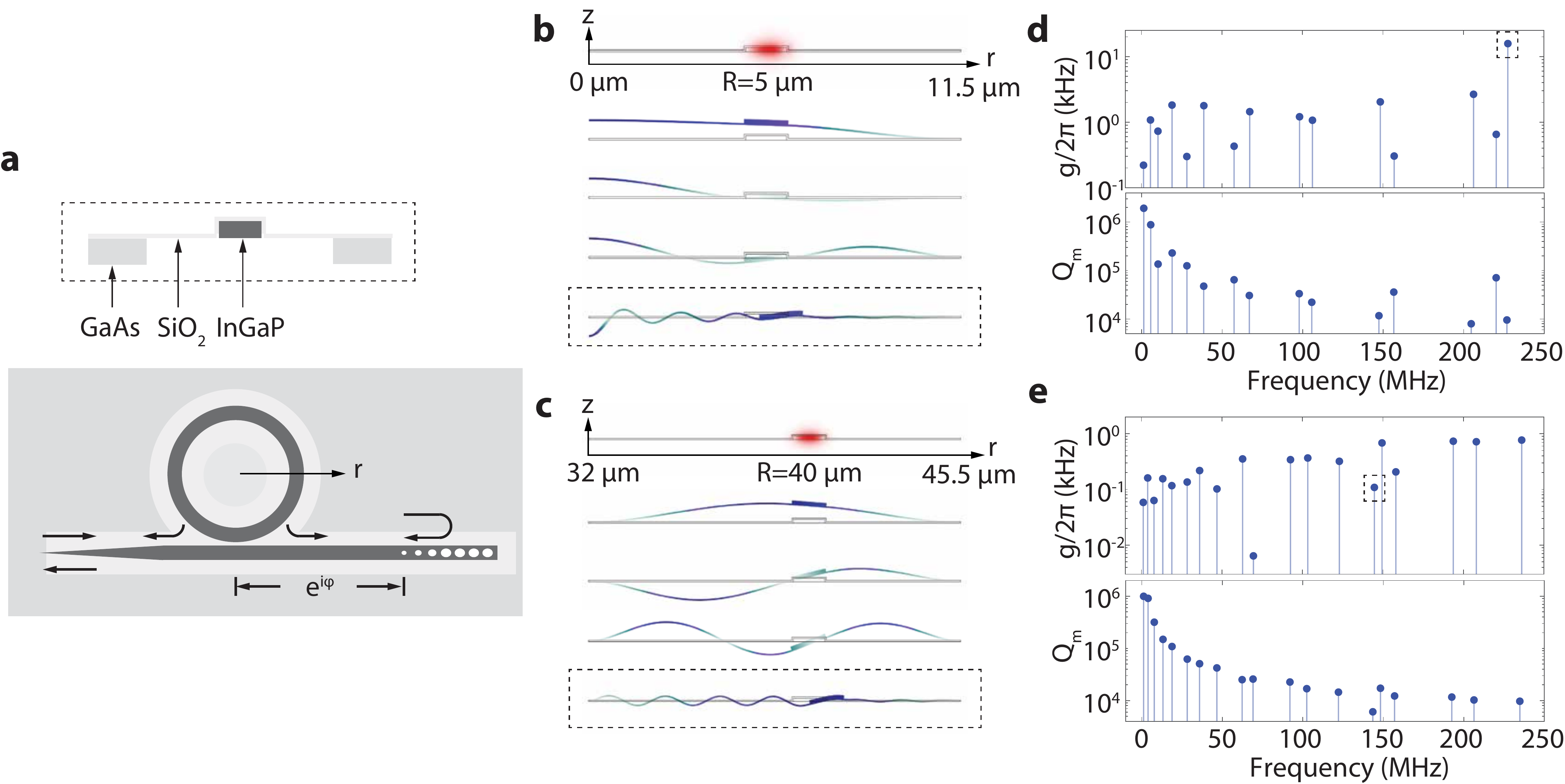}
\caption{\textbf{a}. A schematic plot of the ``cavity-in-a-membrane'' optomechanical system. The photonic circuit consists of a ring resonator side-coupled to a waveguide which is terminated with a photonic crystal mirror and an adiabatic coupler. \textbf{b} and \textbf{c}.  Simulated electric field ($E_r$) of the TE$_{00}$ optical mode and total displacement of a few membrane and ring mechanical modes of the 5-$\mu$m-radius ring (\textbf{b}) and the 40-$\mu$m-radius ring (\textbf{c}). The central membrane of the 5-$\mu$m-radius ring is fully released while that of the 40-$\mu$m-radius ring is partially released with the anchoring positions labeled. \textbf{d} and \textbf{e}. Simulated optomechanical coupling and mechanical quality factor of the 5-$\mu$m-radius ring (\textbf{d}) and the 40-$\mu$m-radius ring (\textbf{e}). The breathing ring mode is highlighted by the dashed box.  
}
\label{fig:modes}
\end{center}
\end{figure*}

\section{Cavity-in-a-membrane optomechanical system}
The ``cavity-in-a-membrane'' optomechanical system is illustrated in Fig. \ref{fig:modes}a. The photonic circuit consisting of a waveguide-coupled ring resonator is embedded in a suspended membrane. The waveguide is terminated with a photonic crystal mirror at one end and an adiabatic coupler at the other. In this study, the photonic circuit and membrane are made from thin-film InGaP (150 nm) and silicon dioxide (50 nm) by atomic layer deposition, respectively (see Appendix \ref{App:fab} for fabrication and an image of the fabricated device). Ultrathin membranes down to a few atomic layers can be fabricated by undercut of the silicon dioxide adjacent to the InGaP circuit. The membrane attached to the ring has an average inner and outer width of 8 $\mu$m and 5.5 $\mu$m, respectively (for rings with a radius less than 8 $\mu$m, the interior region of the ring is thus fully released). We performed finite element simulation of the released ring to identify the mechanical modes. The waveguide is not included in the simulation because the mechanical coupling between the two regions can be approximated as weak perturbations. Two types of mechanical modes, including the flexural membrane mode and the breathing ring mode, can couple with the optical ring resonator. Fig. \ref{fig:modes}b and c show the first few membrane and ring mechanical modes and the fundamental transverse-electric(TE)-like optical resonator mode of the 5-$\mu$m- and 40-$\mu$m-radius rings with a width of $1150$ nm, respectively. The simulated optomechanical coupling, including both photoelastic and moving boundary effects, and mechanical quality factor for the two sizes of ring resonators are plotted in Fig. \ref{fig:modes}d and e, respectively. The simulated mechanical quality factor only involves acoustic radiation and clamping losses.

\begin{figure*}[!htb]
\begin{center}
\includegraphics[width=2\columnwidth]{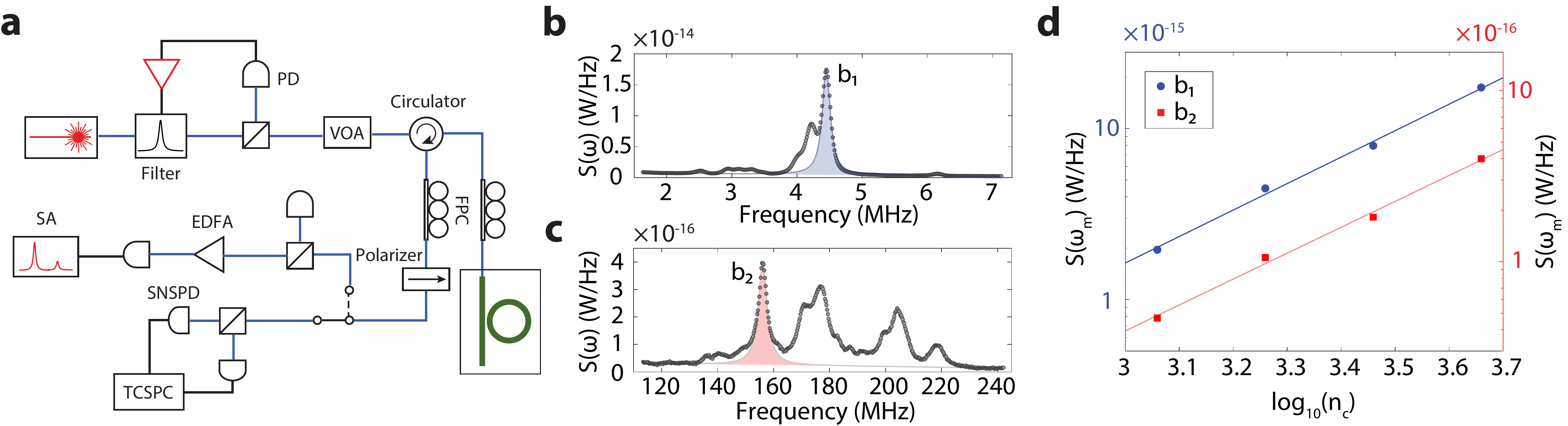}
\caption{\textbf{a}. A schematic plot of the experimental setup for mechanical spectroscopy and photon correlation measurement. FPC, fiber polarization controller. VOA, variable optical attenuator.  EDFA, erbium-doped fiber amplifier. PD, photodetector. SA, spectrum analyzer. SNSPD, superconducting nanowire single-photon detector. TCSPC, time-correlated single-photon counter. \textbf{b} and \textbf{c}. Measured photocurrent noise power spectra showing the flexural membrane modes (\textbf{b}) and breathing ring modes (\textbf{c}) for a 5-$\mu$m-radius ring device. \textbf{d}. Peak noise power versus cavity photon number $n_c$ of the membrane mode $b_1$ and ring mode $b_2$. }
\label{fig:setup}
\end{center}
\end{figure*}

The mechanical spectroscopy was performed at room temperature and ambient environment with the setup illustrated in Fig. \ref{fig:setup}a  (see Appendix \ref{App:measurement} for more details). We describe the result of a device with a $R=5\ \mu$m ring as an example. The loaded quality factor of the TE$_{00}$ 1550 nm-band resonance of the fabricated ring resonator is $Q=1.2\times 10^5$ for the waveguide-ring gap of 250 nm. The intrinsic quality factor of ring resonators with radius ranging from 5 $\mu$m to 80 $\mu$m is similar, which is about $Q_i\approx2.5\times 10^5$ \cite{zhao2022ingap}.  Because the frequency of radiation-pressure coupled mechanical modes is smaller than the optical resonance linewidth, a probe beam blue-detuned from the resonance at the half linewidth is used to drive the optomechanical cavity. An adiabatically tapered fiber-optic coupler is used to transmit light into the device with an efficiency over 70\% \cite{zhao2022ingap}. The reflected light from the device by the photonic crystal mirror is sent to a high-speed photodetector and the noise power spectrum of the photocurrent is measured. Fig. \ref{fig:setup}b and c show the optically-transduced mechanical spectrum, where both the flexural membrane modes and breathing ring modes are observed. There are multiple modes near the breathing ring mode, which might be due to structural mode splitting or the hybridization of the ring mode and higher-order membrane modes. The linewidth of a membrane mode $b_1$ at 4.5 MHz and a ring mode $b_2$ at 156.0 MHz (highlighted in \ref{fig:setup}b and c) is found to be 0.15 MHz and 3.8 MHz, respectively. The dissipation of the mechanical modes is primarily due to gas damping \cite{verbridge2008size}. 
When varying the pump power, the peak noise power of $b_1$ and $b_2$ modes changes linearly with the cavity photon number $n_c$ expectedly (Fig. \ref{fig:setup}d). The optomechanical coupling of $b_1$ and $b_2$ modes is found to be $g/2\pi=1.7$ kHz and $g/2\pi=6.4$ kHz, respectively (Appendix \ref{App:OMcoupling}). The difference from the simulation might be due to mode distortion in the presence of the waveguide and the actual material parameters.

\section{Tunable dissipative filtering}
\begin{figure*}[!htb]
\begin{center}
\includegraphics[width=2\columnwidth]{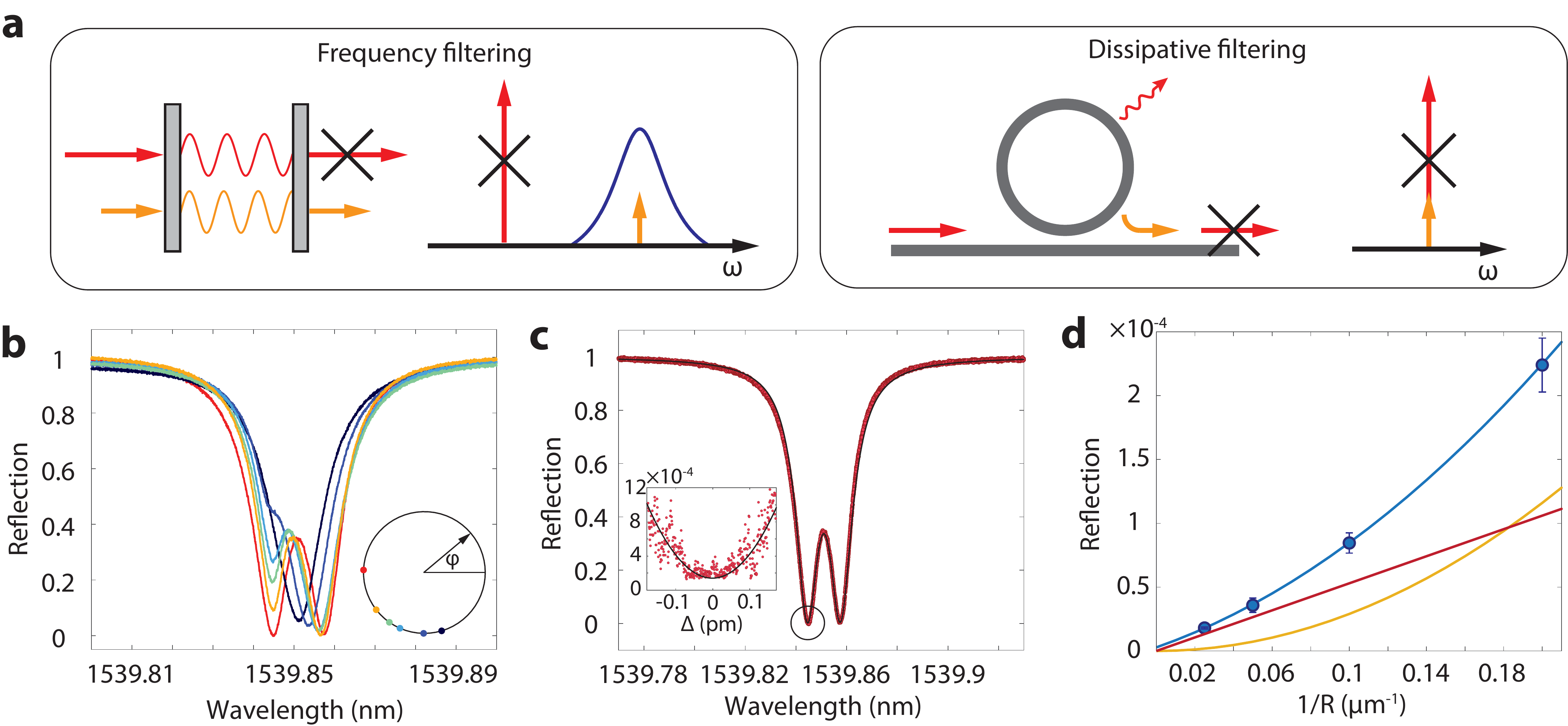}
\caption{\textbf{a}. Schematic plots showing the comparison between the frequency-selective filtering (left) and the dissipative filtering via a one-port cavity-waveguide coupled system (right). Yellow light in the right panel is generated in the cavity. See text. \textbf{b}. Normalized optical reflection spectrum  for different tuning phases (inset), showing the process of tuning one resonance (left) to be critically-coupled (red curve). \textbf{c}. Model fitting of the critically coupled resonance spectrum. Inset shows the reflection near the resonance dip measured by the SNSPD. \textbf{d}. The minimum reflection versus $R^{-1}$. The blue curve is a quadratic fitting of the measured data. The red and yellow curves are the $R^{-1}$ and $R^{-2}$ components corresponding to the mechanically-scattered light and the thermorefractive noise, respectively.} 
\label{fig:filter}
\end{center}
\end{figure*}

In order to detect the photons scattered by the low-frequency mechanical modes via the weak radiation-pressure force, we realize a method to eliminate external pump photons by cavity dissipation engineering, which is distinct from the frequency-selective filtering (Fig. \ref{fig:filter}a). For conventional frequency-selective filters, such as a Fabry-Perot interferometer, light of specific frequencies undergoes constructive or destructive interference leading to transmission or filtering. Dissipative filtering, on the other hand, works for light of different origins and is frequency nonselective. A canonical example is a one-port waveguide-coupled cavity (Fig. \ref{fig:filter}a). When the intrinsic and external loss rates of the cavity are equal, i.e., the critical coupling condition, transmission of light in resonance with the cavity is eliminated, while light originated from the cavity can couple into the waveguide even if its frequency is the same as the input light. However, it is difficult to achieve high-extinction dissipative filtering via fabrication-defined external loss rate as it is sensitive to the waveguide-cavity separation. 

We control the external loss rate of the cavity using a phase-tunable waveguide with extremely fine resolutions (Fig. \ref{fig:modes}a), which works as follows. The fabricated InGaP ring resonator forms standing-wave optical resonances (due to the surface roughness), which couple bi-directionally to the waveguide with an external loss rate $\frac{\kappa_{e0}}{2}$ for each direction. Because the waveguide is terminated with a photonic crystal mirror, the light is reflected back to the input port, leading to an effective one-port cavity-waveguide system. The net external loss rate of the cavity mode thus is approximately given by $\kappa_{e}=\frac{\kappa_{e0}}{2}(1+\sin2\varphi)$, where $\varphi$ is the waveguide propagation phase between the ring and the photonic crystal mirror. As a result, by changing $\varphi$, the cavity external loss rate can be continuously tuned to approach the critical coupling condition, i.e., $\kappa_e=\kappa_i$, when the on-resonance reflection of the external probe from the one-port cavity-waveguide system vanishes. On the other hand, any light generated inside the cavity due to, for example, here mechanical vibration induced Stokes and anti-Stokes scattering is immune to such dissipative cancellation, regardless of the frequency of the scattered light. The output power of the on-resonance scattered light, proportional to $\kappa_e^2/\kappa^4$ (for $\omega_m\ll \kappa$), is in fact maximized under the critical coupling condition.  More rigorous modeling of the dissipative filtering is provided in Appendix \ref{App:device}.

In this study, tuning of the waveguide phase is realized using a tapered optical fiber (taper angle $\approx 2\degree$) in contract with the waveguide top. The introduction of the tuning fiber changes the effective mode index ($n_{\mr{eff}}$) of the waveguide and the incremented propagation phase is given by $\Delta \varphi=2\pi\Delta n_{\mr{eff}}L/\lambda_0$, where $L$ is the length of the fiber on the waveguide. Thus, $\varphi$ can be continuously tuned by changing $L$, which in our experiment is controlled by a motorized stage. For example, according to numerical simulation, for a fiber with 1 $\mu$m radius on top of an InGaP waveguide with dimension of 800 nm$\times$115 nm, an incremented $2\pi$-phase shift can be achieved with $L=37\ \mu$m and a resolution of $8.5\times 10^{-3}$ radians (for 50-nm stage step size). A better resolution can be achieved with a narrower fiber. This method is applied to the $R=5\ \mu$m ring device to tune one of the standing-wave resonance close to the critical coupling condition with a minimum reflection coefficient of $1.4\times 10^{-4}$ (Fig. \ref{fig:filter}b). The fitting of the reflection spectrum near the critical coupling shows an internal quality factor of $Q_i=2.6\times 10^5$ (Fig. \ref{fig:filter}c). 

The minimum reflection achieved via the dissipative filtering reveals a quadratic dependence of the residual light on the inverse of the ring radius as shown in Fig. \ref{fig:filter}d. The minimum reflection of the $R=40\ \mu$m ring is close to $1.8\times 10^{-5}$, indicating a dissipative tuning resolution close to or better than $10^{-5}$. The residual light has two major sources including Stokes/anti-Stokes scattering by the mechanical modes and thermorefractive noise \cite{gorodetsky2004fundamental,schliesser2008high}. The amount of the mechanically scattered light is roughly proportional to $g^2/\kappa^2\propto 1/R$ (Appendix \ref{App:OMcoupling}), because the optomechanical coupling scales with $1/\sqrt{R}$ and ring resonators with different radius have a similar $Q$ \cite{zhao2022ingap}. On the other hand, the power of thermorefractive noise roughly scales as $1/R^2$ (Appendix \ref{App:thermo}). Thus, the $1/R$ and $1/R^2$ component of the minimum reflection corresponds to the amount of the mechanically-scattered light and thermorefractive noise, respectively (Fig. \ref{fig:filter}d).

\begin{figure*}[!htb]
\begin{center}
\includegraphics[width=2\columnwidth]{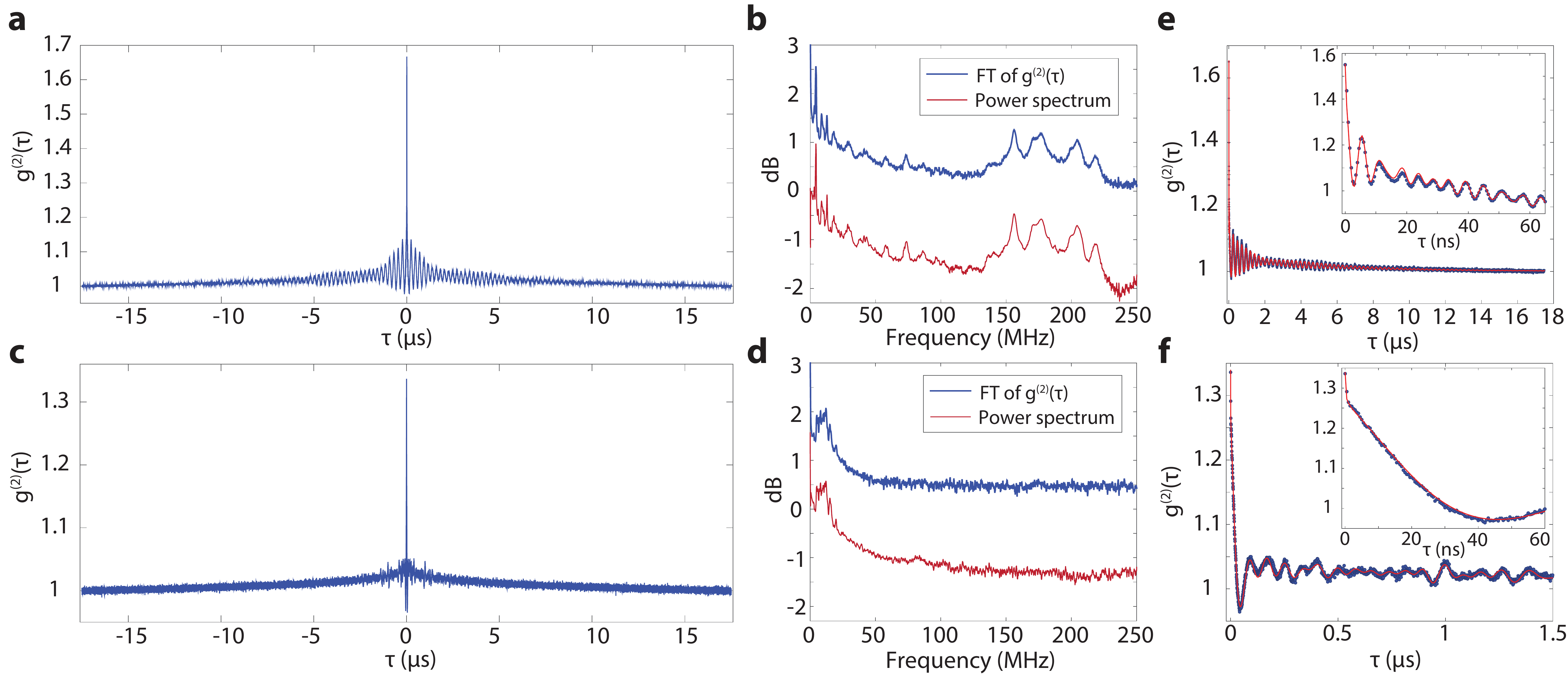}
\caption{\textbf{a}. $g^{(2)}(\tau)$ of a $R=5\ \mu$m ring. \textbf{b}. Fourier transformation (FT) of $g^{(2)}(\tau)$ and noise power spectrum of the $5\ \mu$m ring. The curves are offset for clarity. \textbf{c}. $g^{(2)}(\tau)$ of a $R=40\ \mu$m ring. \textbf{d}. Fourier transformation of $g^{(2)}(\tau)$ and noise power spectrum of the $40\ \mu$m ring. \textbf{e} and \textbf{f}. Model fitting (red) of the $g^{(2)}(\tau)$ of the 5-$\mu$m ring (\textbf{e}) and the 40-$\mu$m ring (\textbf{f}). }
\label{fig:g2}
\end{center}
\end{figure*}

\section{Photon-phonon correlations}

The dissipative attenuation of the probe beam allows measurement of the correlations of the photons scattered by the low-frequency mechanical modes as well as the phonon statistics.  Under a parametric pump, the optomechanical cavity with a low-frequency mechanical mode ($\omega_m\ll \kappa$) is described by a linearized Hamiltonian $\hat{H}_{\mr{OM}}=G(a+a^{\dag})(b+b^{\dag})$, where $a(b)$ is the annihilation operator of the optical(mechanical) mode, $G=g\sqrt{n_c}$ and $n_c$ is the cavity photon number. The output light from the cavity contains the mechanically-scattered photons, which inherit the statistical property of the phonons \cite{cohen2015phonon}, and the residual pump light, i.e., $a_{\mr{out}}(t)=s a_{\mr{in}}(t)+u b_{\mr{in}}(t)+v b^\dagger_{\mr{in}}(t)$, where $s$, $u$ and $v$ are the coefficients of direct reflection, anti-Stokes and Stokes scattering, respectively. As a result, the the self-correlation of the output field contains the cross-correlation between photons and phonons under a continuous pump. For a thermal mechanical state, the normalized second-order correlation function of the output light is given by (Appendix \ref{App:correlation}) 
\begin{widetext}
\begin{equation}
\label{eq:g2model}
\begin{aligned}
g^{(2)}(\tau)=1+\frac{(|u|^2+|v|^2)^2 I_b^2|g_b^{(1)}(\tau)|^2+2|s|^2 (|u|^2+|v|^2) I_a I_b|g_b^{(1)}(\tau)|\cos(\omega_m\tau)+2|uv|^2 I_b^2 |g_b^{(1)}(\tau)|^2\cos(2\omega_m\tau)}{\Big(|s|^2I_a+(|u|^2+|v|^2) I_b \Big)^2} ,
\end{aligned}
\end{equation}
\end{widetext}
where $|g_b^{(1)}(\tau)|=e^{-\gamma\tau/2}$, $I_a=P_{\mr{in}}/\hbar\omega_p$ and $I_b=\gamma \bar n_m/4$ ($\bar n_m$ is the average phonon occupation of the mechanical mode and $\gamma$ is the mechanical dissipation rate) are the input photon and phonon flux, respectively. Because the mechanical modes are in classical states, the correlation function between photons and phonons are factorized. It is seen from Eq. \ref{eq:g2model} that, when $|s|^2\approx |u|^2, |v|^2$, the correlations between photons and phonons could be observed.

We measured the correlation of output photons of the $R=5\ \mu$m ring device with a cavity photon number $n_c\approx 2$. The device is tuned to be critically coupled with a dip reflection of $1.4\times 10^{-4}$ (Fig. \ref{fig:filter}c). The probe beam is frequency locked near the dip with a stable reflection coefficient (Appendix \ref{App:measurement}). When the reflection coefficient is sufficiently small, the second-order correlation function of the output light is manifested with $g^{(2)}(0)>1$ due to the thermal mechanical state as well as oscillations due to the beating between the mechanically-scattered photons and the residual probe photons. Fig. \ref{fig:g2}a shows the measured $g^{(2)}(\tau)$ for reflection coefficient $|s|^2=2.9\times 10^{-4}$. For this measurement, since the residual probe light is at a level higher than the amount of scattered light, the dominant term in Eq. \ref{eq:g2model} is $g_{b}^{(1)}(\tau)\cos(\omega_{m}\tau)$, whose Fourier transform yields the mechanical spectrum \cite{walls2007quantum}. This is shown in Fig. \ref{fig:g2}b, where the Fourier transform of $g^{(2)}(\tau)$ and the noise power spectrum measured by the photodetector are almost identical.  The low-frequency rising slope in the noise power spectrum is due to the thermorefractive noise \cite{gorodetsky2004fundamental,schliesser2008high}.  Fig. \ref{fig:g2}c and d show the measured photon correlation (resonance dip of $1.8\times 10^{-5}$ and locked reflection level of $6.9\times 10^{-5}$) and the corresponding mechanical spectrum of a $R=40\ \mu$m ring, respectively. The measured photon correlation, containing multi-frequency oscillations due to multiple optically-coupled mechanical modes, can be fitted using a multimode  generalization of Eq. \ref{eq:g2model} including the thermorefractive noise (Appendix \ref{App:correlation}), as shown in Fig. \ref{fig:g2}e and f for the $R=5\ \mu$m and $R=40\ \mu$m ring, respectively. From the fitting, the amount of photons scattered by each mechanical mode as well as the thermorefractive noise can be inferred.

\section{Discussion}
The ``cavity-in-a-membrane'' optomechanical architecture enables detection of mechanical motion of ultrathin membranes defined via lithographic fabrication, which might lead to new modalities of sensing and signal transduction. Further studies of the acoustic loss of atomically-thin membranes at low temperature and in the vacuum could be relevant to phonon dynamics in low-dimensional materials. Despite the observed photon correlation being classical, same approach can be applied to probe quantum correlations between photons and phonons when the mechanical oscillator is prepared in a quantum state. The dissipative filtering method, facilitating the detection of scattered light by mechanical oscillators with frequencies in the MHz range and beyond, will also be useful for study of other low-frequency light scattering processes where optical frequency-selective filtering is difficult.

\vspace{2mm}
\noindent\textbf{Acknowledgements}\\ 
We thank Shengyan Liu and Siyuan Wang for the help on modeling and simulation. This work is supported by US National Science Foundation (Grant No. 1944728) and Office of Naval Research (Grant No. N00014-21-1-2136).

%\onecolumngrid

\appendix
\section{Device fabrication}\label{App:fab}
\begin{figure}[!htb]
\begin{center}
\includegraphics[width=\columnwidth]{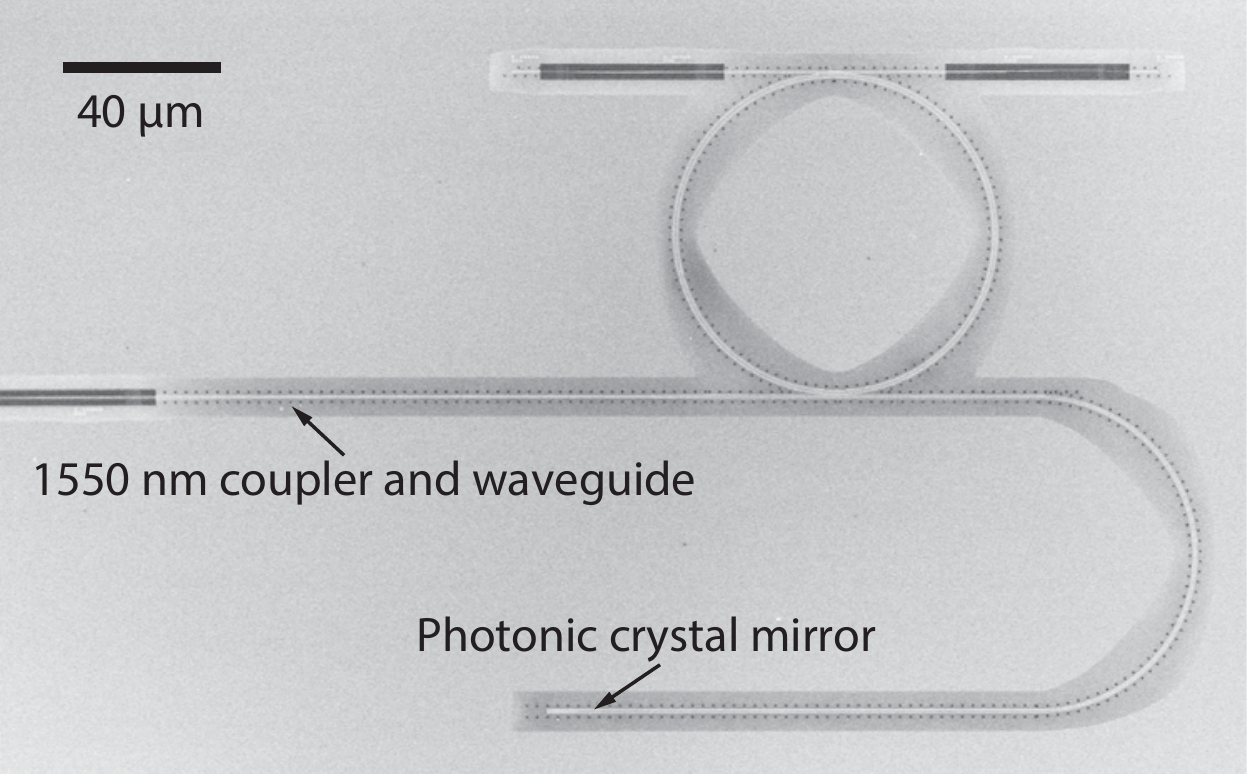}
\caption{Scanning electron microscope image of a device with a 40 $\mu$m radius ring. The top auxiliary waveguide is not used in this work.  }
\label{fig:sem}
\end{center}
\end{figure}

The devices are fabricated from 115 nm thick (measured by scanning electron microscopy) disordered InGaP films grown on GaAs substrate (2 degree off-cut toward the [110]) by metal-organic chemical vapor deposition (T = 545 C, V/III = 48, precursors: trimethylindium, trimethylgallium and PH$_3$). The device pattern is defined using 150 keV electron beam lithography and 150 nm negative tone resist hydrogensilsesquioxane (HSQ). A 20 nm thick layer of silicon dioxide is deposited on InGaP via plasma-enhanced chemical vapor deposition (PECVD) to enhance the adhesion of HSQ. The pattern is transferred to InGaP via inductively coupled plasma reactive-ion etch (ICP-RIE) using Cl$_2$/CH$_4$/Ar gas mixture with a selectivity of InGaP: HSQ: PECVD SiO$_2$ = 240: 90: 80. After a short buffered oxide etch to remove the residual oxide (both HSQ and PECVD oxide), a layer of 50 nm thick silicon dioxide is deposited on the chip via atomic layer deposition. A second electron beam lithography and subsequent ICP-RIE using CHF$_3$ gas are applied to pattern etch-through holes in the silicon dioxide layer for undercut of the InGaP device. The etch-through holes are 200 nm in radius and separated from each other by 2 $\mu$m. Finally, the InGaP device is released from the GaAs substrate using citric acid-based selective etching \cite{uchiyama2006fabrication}, leading to a 11 $\mu$m wide membrane along the waveguide oriented along the [011] and [0$\bar{1}$1] direction. Because the wet etching of GaAs is crystal surface selective, the membrane is anisotropic along the ring with a varying width from 11 to 16 $\mu$m (see Fig. \ref{fig:sem}). The suspended InGaP device is mechanically anchored to the silicon dioxide membrane. See Ref. \cite{zhao2022ingap} for more details.

\section{Device measurement}\label{App:measurement}
A continuous-wave telecom laser (Newport Velocity, TLB-6700) is used to probe the device. The laser beam is filtered by a narrow-bandwidth optical filter (LUNA, 50 GHz FSR, 400 Finesse) to eliminate side-mode emission and broadband amplified spontaneous emission. The filtered light is coupled into the on-chip device via an adiabatic fiber coupler. 
The reflected light from the device passes through a polarization controller and a polarizer to filter the unpolarized and/or orthogonal polarization component of the laser light or scattered light in the fiber. 
The light is then directed either to a high-speed photodetector for measuring the noise power spectral density or a Hanbury-Brown and Twiss setup with two superconducting single-photon detectors (Quantum Opus, 85\% efficiency, 100 Hz dark count rate) for measuring the second-order correlation. For the latter, the reflection signal is locked near the resonance dip by controlling the laser frequency in response to the drift of the resonance due to thermo-optic effect and water vapor in the atmosphere.

\section{Modeling of dissipative filtering }\label{App:device}

\begin{figure}[!htb]
\begin{center}
\includegraphics[width=\columnwidth]{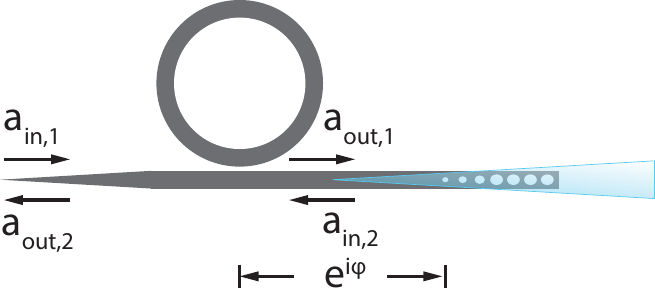}
\caption{Schematic of the input-output relation of the device. A tapered fiber (blue) is used to tune the waveguide phase $\varphi$. }
\label{fig:filtermodel}
\end{center}
\end{figure}

For an ideal ring resonator, the clockwise (cw) and counterclockwise (ccw) resonances are degenerate with frequency $\omega_0$. In fabricated ring resonators,  the clockwise and counterclockwise resonances could couple due to the surface roughness. The input-output equation of the cw and ccw resonances (Fig. \ref{fig:filtermodel}) is given by 
\begin{widetext}
\begin{equation}
\label{eq:eqofmotion}
\begin{aligned}
\frac{d}{dt}\left(\begin{array}{c} a_{\mr{cw}} \\  a_{\mr{ccw}} \end{array}\right) = -i\left(\begin{array}{cc} \omega_0-i\frac{\kappa_{\mr{cw}}}{2} & V\\ V & \omega_0-i\frac{\kappa_{\mr{ccw}}}{2} \end{array}\right)\left(\begin{array}{c} a_{\mr{cw}} \\  a_{\mr{ccw}} \end{array}\right) +\left(\begin{array}{cc} 0 & \sqrt{\kappa_{\mr{cw},e}}\\ \sqrt{\kappa_{\mr{ccw},e}} & 0 \end{array}\right)\left(\begin{array}{c} a_{\mr{in},1} \\  a_{\mr{in},2} \end{array}\right),
\end{aligned}
\end{equation}
\begin{equation}
\label{eq:inout}
\left(\begin{array}{c} a_{\mr{out},1} \\  a_{\mr{out},2} \end{array}\right)=\left(\begin{array}{c} a_{\mr{in},1} \\  a_{\mr{in},2} \end{array}\right)-\left(\begin{array}{cc} 0 & \sqrt{\kappa_{\mr{ccw},e}}\\ \sqrt{\kappa_{\mr{cw},e}} & 0 \end{array}\right)\left(\begin{array}{c} a_{\mr{cw}} \\  a_{\mr{ccw}} \end{array}\right),
\end{equation}
\end{widetext}
where $\kappa_{\mr{cw(ccw)}}=\kappa_{\mr{cw(ccw)},e}+\kappa_{\mr{cw(ccw)},i}$.

The photonic crystal mirror with non-unity reflection acts like a beamsplitter, which relates $a_{\mr{out},1}$ and $a_{\mr{in},2}$ by
\begin{equation}
\label{eq:phc}
a_{\mr{in},2}=e^{i\varphi}(ir e^{i\varphi} a_{\mr{out},1}+t a_{n}),
\end{equation}
where $\varphi$ is the propagation phase between the ring and the photonic crystal mirror, $r$ and $t$ are the real reflection and transmission coefficient of the photonic crystal mirror, and $a_{n}$ represents the noise operator.

By eliminating $a_{\mr{out},1}$ and $a_{\mr{in},2}$ from Eqs.\ref{eq:eqofmotion}-\ref{eq:phc}, we obtain
\begin{widetext}
\begin{equation}
\label{eq:cwccw}
\begin{aligned}
& \frac{d}{dt}\left(\begin{array}{c} a_{\mr{cw}} \\  a_{\mr{ccw}} \end{array}\right)  =-i\left(\begin{array}{cc} \omega_0-i\frac{\kappa_{\mr{cw}}}{2} & V+\sqrt{\kappa_{\mr{cw},e}\kappa_{\mr{ccw},e}} r e^{i2\varphi}\\ V & \omega_0-i\frac{\kappa_{\mr{ccw}}}{2} \end{array}\right)\left(\begin{array}{c} a_{\mr{cw}} \\  a_{\mr{ccw}} \end{array}\right) +\left(\begin{array}{c} i\sqrt{\kappa_{\mr{cw},e}}re^{i2\varphi}\\  \sqrt{\kappa_{\mr{ccw},e}} \end{array}\right) a_{\mr{in},1} +\left(\begin{array}{c} \sqrt{\kappa_{\mr{cw},e}}te^{i\varphi}\\  0 \end{array}\right) a_{n},
\end{aligned}
\end{equation}
\begin{equation}\label{eq:cwccwinout}
a_{\mr{out},2}=ire^{i2\varphi}a_{\mr{in},1}-\sqrt{\kappa_{\mr{cw},e}}a_{\mr{cw}}-i\sqrt{\kappa_{\mr{ccw},e}}r e^{i2\varphi}a_{\mr{ccw}}+t e^{i\varphi}a_{n}.
\end{equation}
\end{widetext}

From now on, we assume the cw and ccw resonances have the same dissipation rate and omit the label. 
To obtain the reflection coefficient of the device, we solve Eqs.\ref{eq:cwccw} and \ref{eq:cwccwinout} in the frequency domain by ignoring the noise operator, which yields
\begin{widetext}
\begin{equation}
\label{eq:inout_rig}
a_{\mr{out},2}[\omega]=i\left[ r e^{i2\varphi}-\kappa_e \left(\begin{array}{cc} 1& ire^{i2\varphi} \end{array}\right) \left(\begin{array}{cc} \omega-\omega_0+i\frac{\kappa}{2} & -V-\kappa_{e}r e^{i2\varphi}\\ -V & \omega-\omega_0+i\frac{\kappa}{2} \end{array}\right)^{-1} \left(\begin{array}{c} ire^{i2\varphi}\\1 \end{array}\right) \right]a_{\mr{in},1}[\omega].
\end{equation}
\end{widetext}
The reflection is zero (i.e., $a_{\mr{out},2}[\omega]=0$) when
\begin{equation}\label{R0c1}
\left(\cos2\varphi+\frac{r\Delta\kappa^2}{2V\kappa_e}\right)^2=1- \frac{r^2\Delta\kappa^2}{\kappa_e^2}
\end{equation}
and
\begin{equation}\label{R0c2}
\begin{aligned}
\omega&=\omega_0\pm\sqrt{V^2+\Delta\kappa^2+ V(\kappa_e/r)\cos2\varphi}\\
&=\omega_0\pm\sqrt{V^2+(\Delta\kappa/2)^2\pm V\sqrt{(\kappa_e/r)^2-\Delta\kappa^2}},
\end{aligned}
\end{equation}
where $\Delta\kappa=\kappa_i-\kappa_e$. 

\subsection{Large splitting limit}
In the large splitting limit, i.e., $V\gg\kappa$, Eq. \ref{eq:cwccw} can be easily diagonalized, leading to 
\begin{widetext}
\begin{equation}
\begin{aligned}
& \frac{d}{dt}\left(\begin{array}{c} a_{\mr{e}} \\  a_{\mr{o}} \end{array}\right)  =-i\left(\begin{array}{cc} \omega_{\mr{e}}-i\frac{\kappa_{\mr{e}} }{2} & \\  & \omega_{\mr{o}}-i\frac{\kappa_{\mr{o}} }{2} \end{array}\right)\left(\begin{array}{c} a_{\mr{e}} \\  a_{\mr{o}} \end{array}\right) +\frac{1}{\sqrt{2}}\left(\begin{array}{c} i\sqrt{\kappa_{e}}re^{i2\varphi}+\sqrt{\kappa_{e}} \\   i\sqrt{\kappa_{e}}re^{i2\varphi}-\sqrt{\kappa_{e}} \end{array}\right) a_{\mr{in},1} +\frac{1}{\sqrt{2}}\left(\begin{array}{c} \sqrt{\kappa_{e}}te^{i\varphi}\\  \sqrt{\kappa_{e}}te^{i\varphi} \end{array}\right) a_{n},
\end{aligned}
\end{equation}
\begin{equation}
a_{\mr{out},2}=ire^{i2\varphi}a_{\mr{in},1}-\frac{1}{\sqrt{2}}(i\sqrt{\kappa_{e}}re^{i2\varphi}+\sqrt{\kappa_{e}} )a_{\mr{e}}-\frac{1}{\sqrt{2}}(-i\sqrt{\kappa_{e}}re^{i2\varphi}+\sqrt{\kappa_{e}} )a_{\mr{o}}+t e^{i\varphi}a_{n},
\end{equation}
\end{widetext}
where $a_{\mr{e(o)}}=\frac{1}{\sqrt{2}}(a_{\mr{cw}}\pm a_{\mr{ccw}})$, $\omega_{\mr{e(o)}}=\omega_0\pm (V+ \frac{1}{2}r\kappa_e\cos2\varphi)$, and $\kappa_{\mr{e(o)}}=\kappa\mp r\kappa_e\sin2\varphi$. As a result, in the large splitting limit, both $a_{\mr{e}}$ and $a_{\mr{o}}$ become unidirectionally coupled resonances. The effective external and intrinsic loss rates of them are
\be
\kappa_{\mr{e(o)},e}=\kappa_e\left(\frac{1+r^2}{2}\mp r\sin2\varphi\right)
\ee
and
\be\label{R0V}
\kappa_{\mr{e(o)},i}=\kappa_i+\frac{t^2}{2}\kappa_e.
\ee
The zero-reflection condition at the large splitting limit is
\begin{equation}\label{R0V}
\sin2\varphi=\pm \frac{r\Delta\kappa}{\kappa_e}
\end{equation}
and
\begin{equation}
\begin{aligned}
\omega &=\omega_0\pm (V+ \frac{1}{2}\frac{\kappa_e}{r}\cos2\varphi)\\&=\omega_0\pm (V\pm \frac{1}{2}\sqrt{(\kappa_e/r)^2-\Delta\kappa^2}), 
\end{aligned}
\end{equation}
which is consistent with Eqs. \ref{R0c1} and \ref{R0c2}. From Eq. \ref{R0V}, it is seen that the zero-reflection condition can be achieved only if $\kappa_i \leqslant (1+\frac{1}{r})\kappa_e$. Note Eq. \ref{R0V} is not the same as $\kappa_{\mr{e(o)},e}=\kappa_{\mr{e(o)},i}$ for $r<1$ because of the Fano interference.

\section{Optically transduced mechanical spectrum}\label{App:OMcoupling}
Consider an optomechanical cavity with the Hamiltonian
\begin{equation}
\hat{H}=\omega_c a^{\dag}a+\omega_m b^{\dag}b+ga^{\dag}a(b+b^{\dag}),
\end{equation}
where $a^\dagger(a)$ and $b^\dagger(b)$ are the creation(annihilation) operator of the optical and mechanical modes, respectively.  The optical mode is subject to a parametric drive with frequency $\omega_p$, and thus the optomechanical Hamiltonian can be linearized in the rotating frame of the pump given by
\begin{equation}
\hat{H}=\Delta a^{\dag}a+\omega_m b^{\dag}b+G(a+a^{\dag})(b+b^{\dag}) ,
\end{equation}
where $\Delta=\omega_c-\omega_p$, $G=g\sqrt{n_c}$, $n_c$ is the cavity photon number, and $a$ corresponds to the quantum fluctuation of the cavity field.
The Heisenberg-Langevin equations of motion for the coupled optical and mechanical modes are given by
\begin{equation}
\dot{a}=(-i\Delta-\frac{\kappa}{2})a-iG(b+b^{\dag})+\sqrt{\kappa_{e}}a_{\mr{in}}+\sqrt{\kappa_{i}}a_{\mr{in},i},
\end{equation}
\begin{equation}
\dot{b}=(-i\omega_m-\frac{\gamma}{2})b-iG(a+a^{\dag})+\sqrt{\gamma}b_{\mr{in}},
\end{equation}
with the input-output relation
\begin{equation}
a_{\mr{out}}=a_{\mr{in}}-\sqrt{\kappa_{e}}a,
\end{equation}
where $a_{\mr{in},i}$ is the vacuum noise from the intrinsic loss channel.

By solving the Heisenberg-Langevin equations in the frequency domain, we obtain
\begin{equation}\label{finout}
a_{\mr{out}}(\omega)=s[\omega]a_{\mr{in}}(\omega)+n_{\mr{opt}}[\omega]a_{\mr{in},i}(\omega)+u[\omega]b_{\mr{in}}(\omega)+v[\omega]b^\dagger_{\mr{in}}(\omega),
\end{equation}
where $a^\dagger_{\mr{in}}(\omega)=(a_{\mr{in}}(-\omega))^\dagger$,
\begin{equation}
\begin{aligned}\label{coef}
& s[\omega]= \frac{(\kappa_{i}-\kappa_{e})/2-i(\omega-\Delta)}{\kappa/2-i(\omega-\Delta)} , \\
& n_{\mr{opt}}[\omega]=-\frac{\sqrt{\kappa_i\kappa_e}}{\kappa/2-i(\omega-\Delta)} ,\\
& u[\omega]=iG \frac{\sqrt{\kappa_{e}}}{\kappa/2-i(\omega-\Delta)}\frac{\sqrt{\gamma}}{\gamma/2-i(\omega-\omega_m)} , \\
& v[\omega]= iG \frac{\sqrt{\kappa_{e}}}{\kappa/2-i(\omega-\Delta)}\frac{\sqrt{\gamma}}{\gamma/2-i(\omega+\omega_m)},
\end{aligned}
\end{equation}
and we have omitted higher order terms $O(G^2/\kappa^2)$ given the weak coupling condition. Note $|s[\omega]|^2+|n_{\mr{opt}}[\omega]|^2=1$.
For the sideband-unresolved case $\omega_m\ll \kappa$, the cavity is driven with a blue-detuned pump at the half linewidth, i.e., $\Delta=-\kappa/2$, and Eq. \ref{coef} is simplified to be
\begin{equation}
\begin{aligned}
& s[\omega]\approx \frac{(\kappa_{i}-\kappa_{e})/2-i\kappa/2}{\kappa/2-i\kappa/2} ,  \\
& n_{\mr{opt}}[\omega] \approx -\frac{\sqrt{\kappa_i\kappa_e}}{\kappa/2-i\kappa/2},\\
& u[\omega]\approx iG \frac{\sqrt{\kappa_{e}}}{\kappa/2-i\kappa/2}\frac{\sqrt{\gamma}}{\gamma/2-i(\omega-\omega_m)} , \\
& v[\omega]\approx  iG \frac{\sqrt{\kappa_{e}}}{\kappa/2-i\kappa/2}\frac{\sqrt{\gamma}}{\gamma/2-i(\omega+\omega_m)} .
\end{aligned}
\end{equation}

The spectral density of the normalized photocurrent due to the beating between the carrier and mechanical sidebands, $
I(\omega)=a_{\mr{out}}(\omega)+a^\dagger_{\mr{out}}(\omega)$, is given by
\begin{widetext}
\begin{equation}
\begin{aligned}
S_{\mr{II}}(\omega)  =&\int d\omega' \langle I^{\dag}(\omega)I(\omega') \rangle \\
=& \int d\omega' \langle s^{*}[-\omega]s[\omega']a_{\mr{in}}^{\dag}(\omega)a_{\mr{in}}(\omega') + s^{*}[-\omega']s[\omega]a_{\mr{in}}(\omega)a_{\mr{in}}^{\dag}(\omega')\\
& + n_{\mr{opt}}^{*}[-\omega]n_{\mr{opt}}[\omega']a_{\mr{in},i}^{\dag}(\omega)a_{\mr{in},i}(\omega') + n_{\mr{opt}}^{*}[-\omega']n_{\mr{opt}}[\omega]a_{\mr{in},i}(\omega)a_{\mr{in},i}^{\dag}(\omega')\\
 &+ (u^*[-\omega]+v[\omega])(u[\omega']+v^*[-\omega'])b_{\mr{in}}^{\dag}(\omega)b_{\mr{in}}(\omega')+ (u^*[-\omega']+v[\omega'])(u[\omega]+v^*[-\omega])b_{\mr{in}}(\omega)b_{\mr{in}}^{\dag}(\omega') \rangle\\
 = &|s[\omega]|^2+|n_{\mr{opt}}[\omega]|^2+\frac{4\kappa_e|G|^2}{\kappa^2}\Bigg[\frac{\gamma \bar{n}_m}{(\gamma/2)^2+(\omega+\omega_m)^2}+\frac{\gamma (\bar{n}_m+1)}{(\gamma/2)^2+(\omega-\omega_m)^2} \Bigg]\\
 = &1+\frac{8\kappa_e|G|^2}{\kappa^2}\bar S_m,
\end{aligned}
\end{equation}
\end{widetext}
where $\bar{n}_m=\frac{k_BT}{\hbar\omega_m}$ is the average phonon occupation number and we have used $[a_{\mr{in}}(\omega),a_{\mr{in}}^{\dag}(\omega')]=[a_{\mr{in},i}(\omega),a_{\mr{in},i}^{\dag}(\omega')]=[b_{\mr{in}}(\omega),b_{\mr{in}}^{\dag}(\omega')]=\delta(\omega+\omega')$, $\langle a_{\mr{in}}^{\dag}(\omega')a_{\mr{in}}(\omega)\rangle=\langle a_{\mr{in},i}^{\dag}(\omega')a_{\mr{in},i}(\omega)\rangle=0$, $\langle b_{\mr{in}}^{\dag}(\omega')b_{\mr{in}}(\omega)\rangle=\bar n_m\delta(\omega+\omega') $. 
Thus, the measured noise power spectral density is
\begin{equation}
\label{eq:npsd}
\begin{aligned}
S(\omega) =S_e+\frac{G_e^2}{R_I}\Bigg[S_{\mr{EDFA}}+G_{\mr{EDFA}}^2 S_{\mr{SN}}^2\Bigg(1+\eta\frac{8\kappa_e|G|^2}{\kappa^2}\bar S_m\Bigg)\Bigg],
\end{aligned}
\end{equation}
where $S_e$ is the electronic noise of the detector, $S_{\mr{EDFA}}$ is the noise induced by EDFA, $S_{\mr{SN}}=\sqrt{2P_{\mr{out}}\hbar\omega_o}$ is the shot noise of the light into the detector, $G_e=900\mr{V/W}$ is the conversion gain of the detector, $R_I=50\Omega$ is the impedance of the spectrum analyzer, and $G_{\mr{EDFA}}$ is the EDFA gain.  $\eta=\eta_c\eta_f\eta_d=0.41$ is the total efficiency of the setup, including $\eta_c=0.70$, $\eta_f=0.86$, and $\eta_d=0.68$ for the fiber-optic coupler efficiency, the fiber transmission efficiency from the chip to the detector, and the detector efficiency, respectively.
To characterize the EDFA gain, we measured the input ($P_{\mr{in}}$) and output ($P_{\mr{out}}$) power of the EDFA and the ratio ($\zeta$) of coherent laser power to amplified spontaneous emission power using an optical spectrum analyzer. The EDFA gain is thus given by $G_{\mr{EDFA}}=\zeta P_{\mr{out}}/P_{\mr{in}}$. 

The optomechanical coupling $g$ is inferred from the mechanical resonance noise peak above the noise floor, given by $32 \frac{G_e^2}{R_I} G_{\mr{EDFA}}^2 S_{\mr{SN}}^2 \eta \frac{\kappa_e|G|^2\bar{n}_m}{\kappa^2\gamma}$ according to Eq. \ref{eq:npsd}.

\section{Photon-phonon correlations}\label{App:correlation}
According to Eqs. \ref{finout} and \ref{coef}, the temporal input-output operator relation for a monochromatic input is given by
%\begin{equation}
%a_{\mr{out}}(t)=s[0]a_{\mr{in}}(t)+t[0]a^\dagger_{\mr{in}}(t)+u[0]b_{\mr{in}}(t)e^{-i(\omega_p-\omega_m)t}+v[0]b^\dagger_{\mr{in}}(t)e^{i(\omega_p-\omega_m)t}.
%\end{equation}
\begin{equation}
a_{\mr{out}}(t)=sa_{\mr{in}}(t)+ub_{\mr{in}}(t)+vb^\dagger_{\mr{in}}(t),
\end{equation}
where 
\begin{equation}
\begin{aligned}
& s= \frac{(\kappa_{i}-\kappa_{e})/2+i\Delta}{\kappa/2+i\Delta} , \\
& u=iG \frac{\sqrt{\kappa_{e}}}{\kappa/2+i\Delta}\frac{2}{\sqrt{\gamma}} , \\
& v= iG \frac{\sqrt{\kappa_{e}}}{\kappa/2+i\Delta}\frac{2}{\sqrt{\gamma}}  .
\end{aligned}
\end{equation}
Note the $a$ operator here corresponds to the total optical field and thus is dominated by the coherent pump.
We have omitted the vacuum noise of the intrinsic loss channel because the correlation function is normally ordered. 
Using this, the second-order correlation function of the output photons can be computed as
\begin{widetext}
\begin{equation}
\begin{aligned}
&\langle a^{\dag}_{\mr{out}}(0)a^{\dag}_{\mr{out}}(\tau)a_{\mr{out}}(\tau)a_{\mr{out}}(0)\rangle\\
= &  |s|^4 \langle a^{\dag}_{\mr{in}}(0)a^{\dag}_{\mr{in}}(\tau)a_{\mr{in}}(\tau)a_{\mr{in}}(0)\rangle +(|u|^2+|v|^2)^2 \langle b^{\dag}_{\mr{in}}(0)b^{\dag}_{\mr{in}}(\tau)b_{\mr{in}}(\tau)b_{\mr{in}}(0)\rangle \\
&      + |su|^2 \left(\langle a^{\dag}_{\mr{in}}(0)a_{\mr{in}}(\tau)b_{\mr{in}}^{\dag}(\tau)b_{\mr{in}}(0)\rangle +\langle a^{\dag}_{\mr{in}}(\tau)a_{\mr{in}}(0)b_{\mr{in}}^{\dag}(0)b_{\mr{in}}(\tau)\rangle  \right) \\
& + |sv|^2 \left( \langle a^{\dag}_{\mr{in}}(0)a_{\mr{in}}(\tau)b_{\mr{in}}^{\dag}(0)b_{\mr{in}}(\tau)\rangle +\langle a^{\dag}_{\mr{in}}(\tau)a_{\mr{in}}(0)b_{\mr{in}}^{\dag}(\tau)b_{\mr{in}}(0) \rangle  \right) \\
& + |uv|^2 \left(\langle  b^{\dag}_{\mr{in}}(0)b^{\dag}_{\mr{in}}(0)b_{\mr{in}}(\tau)b_{\mr{in}}(\tau)\rangle  +\langle  b^{\dag}_{\mr{in}}(\tau)b^{\dag}_{\mr{in}}(\tau)b_{\mr{in}}(0)b_{\mr{in}}(0)\rangle  \right)\\
& + |s|^2(|u|^2+|v|^2) \left( \langle  a^{\dag}_{\mr{in}}(0)a_{\mr{in}}(0)b_{\mr{in}}^{\dag}(\tau)b_{\mr{in}}(\tau)\rangle +\langle a^{\dag}_{\mr{in}}(\tau)a_{\mr{in}}(\tau)b_{\mr{in}}^{\dag}(0)b_{\mr{in}}(0) \rangle  \right)  \\
=&|s|^4 G^{(2)}_a(\tau)+ (|u|^2+|v|^2)^2 (G^{(1)}_b(0)^2+|G^{(1)}_b(\tau)|^2)+2|s|^2(|u|^2+|v|^2) |G^{(1)}_a(\tau)G^{(1)}_b(\tau)|\cos(\omega_m\tau)  \\
& +2|uv|^2 |G^{(1)}_b(\tau)|^2\cos(2\omega_m\tau)+2|s|^2(|u|^2+|v|^2) G^{(1)}_a(0)G^{(1)}_b(0),
\end{aligned}
\end{equation}
\end{widetext}
where we have used Wick's theorem for calculation of the second-order correlation function of the thermal mechanical state. For thermal mechanical state, $G^{(1)}_b(\tau)=G^{(1)}_b(0)e^{-\gamma\tau/2}$.

The normalized second-order correlation function thus is 
\begin{widetext}
\begin{equation}\label{g2}
\begin{aligned}
g^{(2)}(\tau)&=\frac{\langle a^{\dag}_{\mr{out}}(0)a^{\dag}_{\mr{out}}(\tau)a_{\mr{out}}(\tau)a_{\mr{out}}(0)\rangle }{\langle a^{\dag}_{\mr{out}}(0)a_{\mr{out}}(0)\rangle^2}\\
& =1+\frac{(|u|^2+|v|^2)^2 I_b^2|g_b^{(1)}(\tau)|^2+2|s|^2 (|u|^2+|v|^2) I_a I_b|g_b^{(1)}(\tau)|\cos(\omega_m\tau)+2|uv|^2 I_b^2 |g_b^{(1)}(\tau)|^2\cos(2\omega_m\tau)}{\Big(|s|^2I_a+(|u|^2+|v|^2) I_b \Big)^2},
\end{aligned}
\end{equation}
\end{widetext}
where $I_a=P_{\mr{in}}/\hbar\omega_p$ and $I_b=\gamma \bar n_m/4$ are    the input photon and phonon flux, respectively. Note $g^{(2)}(\tau)$ is independent of input optical power, because $u^2, v^2\propto G^2\propto I_a$.

When there are multiple optically-coupled mechanical modes, the normalized second-order correlation function of the output photon is given by
%\begin{equation}
%\label{eq:eqofmotion}
%\begin{aligned}
%g^{(2)}(\tau)=1+\frac{\sum\limits_{i} |r_i|^4 N_{b_i}^2 |g_{b_i}^{(1)}(\tau)|^2+\sum\limits_i 2|sr_i|^2 N_a N_{b_i} |g_{b_i}^{(1)}(\tau)|\cos(\omega_{b_i}\tau)+\sum\limits_{ij} 2|r_i r_j|^2 N_{b_i} N_{b_j} |g_{b_i}^{(1)}(\tau)g_{b_j}^{(1)}(\tau)|\cos((\omega_{b_i}-\omega_{b_j})\tau)}{(|s|^2 N_a+\sum\limits_i |r_i|^2 N_{b_i})^2} .
%\end{aligned}
%\end{equation}

\begin{widetext}
\begin{equation}
\begin{aligned}
g^{(2)}(\tau) =1+\Big( & \sum\limits_{i}(|u_i|^2+|v_i|^2)^2 I_{b_i}^2|g_{b_i}^{(1)}(\tau)|^2+2|s|^2 (|u_i|^2+|v_i|^2) I_a I_{b_i}|g_{b_i}^{(1)}(\tau)|\cos(\omega_{m_i}\tau)+2|u_iv_i|^2 I_{b_i}^2 |g_{b_i}^{(1)}(\tau)|^2 \cos(2\omega_{m_i}\tau)  \\
& + \sum\limits_{ij} 2|u_iv_j^*+u_jv_i^*|^2  I_{b_i}I_{b_j} |g_{b_i}^{(1)}(\tau)g_{b_j}^{(1)}(\tau)|\left(\cos(\omega_{m_i}-\omega_{m_j})\tau+ \cos(\omega_{m_i}+\omega_{m_j})\tau\right) \Big) \\
&\Big/\Big(|s|^2I_a+\sum\limits_{i}(|u_i|^2+|v_i|^2) I_{b_i} \Big)^2 .
\end{aligned}
\end{equation}
\end{widetext}

The dominant modes in the measured mechanical spectrum are used for the fitting of the measured $g^{(2)}(\tau)$, which are 9 and 11 modes for the 5-$\mu$m- and 40-$\mu$m-radius ring, respectively. The frequencies of the mechanical modes are inferred from the spectrum directly, while $s^2I_a$, $u_i^2 I_{b_i}$ and $v_i^2 I_{b_i}$ and the linewidth of the mechanical modes  are treated as fitting parameters.

\section{Thermorefractive noise}\label{App:thermo}
Thermorefractive effect is the fluctuation of refractive index due to the fluctuation of temperature in a finite volume. It will cause jitter of the optical cavity frequency and generation of random photons when the cavity is illuminated. The power spectral density of thermorefractive noise was computed and measured for microsphere and micro-toroid cavities before \cite{gorodetsky2004fundamental,schliesser2008high}. It was shown that the thermorefractive cavity frequency jitter, $\delta\omega_c$, of the whispering gallery mode is proportional to $1/R$, where $R$ is the cavity radius. The ring optical mode is in good approximation with those whispering gallery mode and we assume $\delta\omega_c$ have the same scaling. 
Under an external pump, the cavity field amplitude is given by 
$a[\omega]=\frac{\sqrt{\kappa_e}}{i(\omega-\omega_c)+\kappa/2}a_{\mr{in}}$.
The thermorefractive noise amplitude under resonant driving caused by resonance frequency jitter $\delta\omega_c$ thus is $\delta a[\omega_c]= i\frac{4\sqrt{\kappa_e}}{\kappa^2}\delta\omega_c a_{\mr{in}}$ and the output noise power is $|\delta a_{\mr{out}}|^2=\kappa_e|\delta a|^2\propto \frac{1}{R^2}$, given similar $\kappa$ for cavities with different radii.

The second-order correlation function after incorporating the thermorefractive noise is given by
\begin{widetext}
\begin{equation}
\begin{aligned}
g^{(2)}(\tau)&
& =1+\frac{\mr{photon\mbox{-}phonon}+2|s|^2 |p|^2 I_a^2|g_\mr{th}^{(1)}(\tau)|+ |p|^4I_a^2|g_\mr{th}^{(1)}(\tau)|^2}{\Big(\mathrm{photon\mbox{-}phonon}+|p|^2I_a\Big)^2},
\end{aligned}
\end{equation}
\end{widetext}
where $p$ is the effective scattering coefficient of the thermorefractive noise, 
$|g_{\mr{th}}^{(1)}(\tau)|=e^{-(\tau/\tau_{\mr{th}})^{n}}$ with $n=1/3$, and photon-phonon terms are given in Eq. \ref{g2}. Both $|p|$ and $\tau_{\mr{th}}$ are fitting parameters for $g^{(2)}(\tau)$. The form of $g_{\mr{th}}^{(1)}(\tau)$ is inferred from silicon-on-insulator (SOI) ring resonators operated at the similar condition (i.e, critically coupled using dissipative filtering). Because of the structural rigidity, SOI ring resonators are free of low-frequency vibrational modes and the observed $g^{(2)}(\tau)>1$ at low reflection is purely due to the correlation of the thermorefractive noise. Given the same type of ring resonators, the correlation function of thermorefractive noise in InGaP devices will be the same as the SOI device but with a different $\tau_{\mr{th}}$ which depends on material parameters and device dimensions.

%\bibliographystyle{naturemag}
%\bibliography{./reference}

\begin{thebibliography}{21}%
\makeatletter
\providecommand \@ifxundefined [1]{%
 \@ifx{#1\undefined}
}%
\providecommand \@ifnum [1]{%
 \ifnum #1\expandafter \@firstoftwo
 \else \expandafter \@secondoftwo
 \fi
}%
\providecommand \@ifx [1]{%
 \ifx #1\expandafter \@firstoftwo
 \else \expandafter \@secondoftwo
 \fi
}%
\providecommand \natexlab [1]{#1}%
\providecommand \enquote  [1]{``#1''}%
\providecommand \bibnamefont  [1]{#1}%
\providecommand \bibfnamefont [1]{#1}%
\providecommand \citenamefont [1]{#1}%
\providecommand \href@noop [0]{\@secondoftwo}%
\providecommand \href [0]{\begingroup \@sanitize@url \@href}%
\providecommand \@href[1]{\@@startlink{#1}\@@href}%
\providecommand \@@href[1]{\endgroup#1\@@endlink}%
\providecommand \@sanitize@url [0]{\catcode `\\12\catcode `\$12\catcode
  `\&12\catcode `\#12\catcode `\^12\catcode `\_12\catcode `\%12\relax}%
\providecommand \@@startlink[1]{}%
\providecommand \@@endlink[0]{}%
\providecommand \url  [0]{\begingroup\@sanitize@url \@url }%
\providecommand \@url [1]{\endgroup\@href {#1}{\urlprefix }}%
\providecommand \urlprefix  [0]{URL }%
\providecommand \Eprint [0]{\href }%
\providecommand \doibase [0]{http://dx.doi.org/}%
\providecommand \selectlanguage [0]{\@gobble}%
\providecommand \bibinfo  [0]{\@secondoftwo}%
\providecommand \bibfield  [0]{\@secondoftwo}%
\providecommand \translation [1]{[#1]}%
\providecommand \BibitemOpen [0]{}%
\providecommand \bibitemStop [0]{}%
\providecommand \bibitemNoStop [0]{.\EOS\space}%
\providecommand \EOS [0]{\spacefactor3000\relax}%
\providecommand \BibitemShut  [1]{\csname bibitem#1\endcsname}%
\let\auto@bib@innerbib\@empty
%</preamble>
\bibitem [{\citenamefont {Aspelmeyer}\ \emph {et~al.}(2014)\citenamefont
  {Aspelmeyer}, \citenamefont {Kippenberg},\ and\ \citenamefont
  {Marquardt}}]{aspelmeyer2014cavity}%
  \BibitemOpen
  \bibfield  {author} {\bibinfo {author} {\bibfnamefont {Markus}\ \bibnamefont
  {Aspelmeyer}}, \bibinfo {author} {\bibfnamefont {Tobias~J}\ \bibnamefont
  {Kippenberg}}, \ and\ \bibinfo {author} {\bibfnamefont {Florian}\
  \bibnamefont {Marquardt}},\ }\bibfield  {title} {\enquote {\bibinfo {title}
  {Cavity optomechanics},}\ }\href@noop {} {\bibfield  {journal} {\bibinfo
  {journal} {Reviews of Modern Physics}\ }\textbf {\bibinfo {volume} {86}},\
  \bibinfo {pages} {1391} (\bibinfo {year} {2014})}\BibitemShut {NoStop}%
\bibitem [{\citenamefont {Cohen}\ \emph {et~al.}(2015)\citenamefont {Cohen},
  \citenamefont {Meenehan}, \citenamefont {MacCabe}, \citenamefont
  {Gr{\"o}blacher}, \citenamefont {Safavi-Naeini}, \citenamefont {Marsili},
  \citenamefont {Shaw},\ and\ \citenamefont {Painter}}]{cohen2015phonon}%
  \BibitemOpen
  \bibfield  {author} {\bibinfo {author} {\bibfnamefont {Justin~D}\
  \bibnamefont {Cohen}}, \bibinfo {author} {\bibfnamefont {Se{\'a}n~M}\
  \bibnamefont {Meenehan}}, \bibinfo {author} {\bibfnamefont {Gregory~S}\
  \bibnamefont {MacCabe}}, \bibinfo {author} {\bibfnamefont {Simon}\
  \bibnamefont {Gr{\"o}blacher}}, \bibinfo {author} {\bibfnamefont {Amir~H}\
  \bibnamefont {Safavi-Naeini}}, \bibinfo {author} {\bibfnamefont {Francesco}\
  \bibnamefont {Marsili}}, \bibinfo {author} {\bibfnamefont {Matthew~D}\
  \bibnamefont {Shaw}}, \ and\ \bibinfo {author} {\bibfnamefont {Oskar}\
  \bibnamefont {Painter}},\ }\bibfield  {title} {\enquote {\bibinfo {title}
  {Phonon counting and intensity interferometry of a nanomechanical
  resonator},}\ }\href@noop {} {\bibfield  {journal} {\bibinfo  {journal}
  {Nature}\ }\textbf {\bibinfo {volume} {520}},\ \bibinfo {pages} {522--525}
  (\bibinfo {year} {2015})}\BibitemShut {NoStop}%
\bibitem [{\citenamefont {Meenehan}\ \emph {et~al.}(2015)\citenamefont
  {Meenehan}, \citenamefont {Cohen}, \citenamefont {MacCabe}, \citenamefont
  {Marsili}, \citenamefont {Shaw},\ and\ \citenamefont
  {Painter}}]{meenehan2015pulsed}%
  \BibitemOpen
  \bibfield  {author} {\bibinfo {author} {\bibfnamefont {Se{\'a}n~M}\
  \bibnamefont {Meenehan}}, \bibinfo {author} {\bibfnamefont {Justin~D}\
  \bibnamefont {Cohen}}, \bibinfo {author} {\bibfnamefont {Gregory~S}\
  \bibnamefont {MacCabe}}, \bibinfo {author} {\bibfnamefont {Francesco}\
  \bibnamefont {Marsili}}, \bibinfo {author} {\bibfnamefont {Matthew~D}\
  \bibnamefont {Shaw}}, \ and\ \bibinfo {author} {\bibfnamefont {Oskar}\
  \bibnamefont {Painter}},\ }\bibfield  {title} {\enquote {\bibinfo {title}
  {Pulsed excitation dynamics of an optomechanical crystal resonator near its
  quantum ground state of motion},}\ }\href@noop {} {\bibfield  {journal}
  {\bibinfo  {journal} {Physical Review X}\ }\textbf {\bibinfo {volume} {5}},\
  \bibinfo {pages} {041002} (\bibinfo {year} {2015})}\BibitemShut {NoStop}%
\bibitem [{\citenamefont {Riedinger}\ \emph {et~al.}(2016)\citenamefont
  {Riedinger}, \citenamefont {Hong}, \citenamefont {Norte}, \citenamefont
  {Slater}, \citenamefont {Shang}, \citenamefont {Krause}, \citenamefont
  {Anant}, \citenamefont {Aspelmeyer},\ and\ \citenamefont
  {Gr{\"o}blacher}}]{riedinger2016non}%
  \BibitemOpen
  \bibfield  {author} {\bibinfo {author} {\bibfnamefont {Ralf}\ \bibnamefont
  {Riedinger}}, \bibinfo {author} {\bibfnamefont {Sungkun}\ \bibnamefont
  {Hong}}, \bibinfo {author} {\bibfnamefont {Richard~A}\ \bibnamefont {Norte}},
  \bibinfo {author} {\bibfnamefont {Joshua~A}\ \bibnamefont {Slater}}, \bibinfo
  {author} {\bibfnamefont {Juying}\ \bibnamefont {Shang}}, \bibinfo {author}
  {\bibfnamefont {Alexander~G}\ \bibnamefont {Krause}}, \bibinfo {author}
  {\bibfnamefont {Vikas}\ \bibnamefont {Anant}}, \bibinfo {author}
  {\bibfnamefont {Markus}\ \bibnamefont {Aspelmeyer}}, \ and\ \bibinfo {author}
  {\bibfnamefont {Simon}\ \bibnamefont {Gr{\"o}blacher}},\ }\bibfield  {title}
  {\enquote {\bibinfo {title} {Non-classical correlations between single
  photons and phonons from a mechanical oscillator},}\ }\href@noop {}
  {\bibfield  {journal} {\bibinfo  {journal} {Nature}\ }\textbf {\bibinfo
  {volume} {530}},\ \bibinfo {pages} {313--316} (\bibinfo {year}
  {2016})}\BibitemShut {NoStop}%
\bibitem [{\citenamefont {Wallucks}\ \emph {et~al.}(2020)\citenamefont
  {Wallucks}, \citenamefont {Marinkovi{\'c}}, \citenamefont {Hensen},
  \citenamefont {Stockill},\ and\ \citenamefont
  {Gr{\"o}blacher}}]{wallucks2020quantum}%
  \BibitemOpen
  \bibfield  {author} {\bibinfo {author} {\bibfnamefont {Andreas}\ \bibnamefont
  {Wallucks}}, \bibinfo {author} {\bibfnamefont {Igor}\ \bibnamefont
  {Marinkovi{\'c}}}, \bibinfo {author} {\bibfnamefont {Bas}\ \bibnamefont
  {Hensen}}, \bibinfo {author} {\bibfnamefont {Robert}\ \bibnamefont
  {Stockill}}, \ and\ \bibinfo {author} {\bibfnamefont {Simon}\ \bibnamefont
  {Gr{\"o}blacher}},\ }\bibfield  {title} {\enquote {\bibinfo {title} {A
  quantum memory at telecom wavelengths},}\ }\href@noop {} {\bibfield
  {journal} {\bibinfo  {journal} {Nature Physics}\ }\textbf {\bibinfo {volume}
  {16}},\ \bibinfo {pages} {772--777} (\bibinfo {year} {2020})}\BibitemShut
  {NoStop}%
\bibitem [{\citenamefont {Fiaschi}\ \emph {et~al.}(2021)\citenamefont
  {Fiaschi}, \citenamefont {Hensen}, \citenamefont {Wallucks}, \citenamefont
  {Benevides}, \citenamefont {Li}, \citenamefont {Alegre},\ and\ \citenamefont
  {Gr{\"o}blacher}}]{fiaschi2021optomechanical}%
  \BibitemOpen
  \bibfield  {author} {\bibinfo {author} {\bibfnamefont {Niccol{\`o}}\
  \bibnamefont {Fiaschi}}, \bibinfo {author} {\bibfnamefont {Bas}\ \bibnamefont
  {Hensen}}, \bibinfo {author} {\bibfnamefont {Andreas}\ \bibnamefont
  {Wallucks}}, \bibinfo {author} {\bibfnamefont {Rodrigo}\ \bibnamefont
  {Benevides}}, \bibinfo {author} {\bibfnamefont {Jie}\ \bibnamefont {Li}},
  \bibinfo {author} {\bibfnamefont {Thiago P~Mayer}\ \bibnamefont {Alegre}}, \
  and\ \bibinfo {author} {\bibfnamefont {Simon}\ \bibnamefont
  {Gr{\"o}blacher}},\ }\bibfield  {title} {\enquote {\bibinfo {title}
  {Optomechanical quantum teleportation},}\ }\href@noop {} {\bibfield
  {journal} {\bibinfo  {journal} {Nature Photonics}\ }\textbf {\bibinfo
  {volume} {15}},\ \bibinfo {pages} {817--821} (\bibinfo {year}
  {2021})}\BibitemShut {NoStop}%
\bibitem [{\citenamefont {Reinhardt}\ \emph {et~al.}(2016)\citenamefont
  {Reinhardt}, \citenamefont {M{\"u}ller}, \citenamefont {Bourassa},\ and\
  \citenamefont {Sankey}}]{reinhardt2016ultralow}%
  \BibitemOpen
  \bibfield  {author} {\bibinfo {author} {\bibfnamefont {Christoph}\
  \bibnamefont {Reinhardt}}, \bibinfo {author} {\bibfnamefont {Tina}\
  \bibnamefont {M{\"u}ller}}, \bibinfo {author} {\bibfnamefont {Alexandre}\
  \bibnamefont {Bourassa}}, \ and\ \bibinfo {author} {\bibfnamefont {Jack~C}\
  \bibnamefont {Sankey}},\ }\bibfield  {title} {\enquote {\bibinfo {title}
  {Ultralow-noise sin trampoline resonators for sensing and optomechanics},}\
  }\href@noop {} {\bibfield  {journal} {\bibinfo  {journal} {Physical Review
  X}\ }\textbf {\bibinfo {volume} {6}},\ \bibinfo {pages} {021001} (\bibinfo
  {year} {2016})}\BibitemShut {NoStop}%
\bibitem [{\citenamefont {Norte}\ \emph {et~al.}(2016)\citenamefont {Norte},
  \citenamefont {Moura},\ and\ \citenamefont
  {Gr{\"o}blacher}}]{norte2016mechanical}%
  \BibitemOpen
  \bibfield  {author} {\bibinfo {author} {\bibfnamefont {Richard~A}\
  \bibnamefont {Norte}}, \bibinfo {author} {\bibfnamefont {Joao~P}\
  \bibnamefont {Moura}}, \ and\ \bibinfo {author} {\bibfnamefont {Simon}\
  \bibnamefont {Gr{\"o}blacher}},\ }\bibfield  {title} {\enquote {\bibinfo
  {title} {Mechanical resonators for quantum optomechanics experiments at room
  temperature},}\ }\href@noop {} {\bibfield  {journal} {\bibinfo  {journal}
  {Physical review letters}\ }\textbf {\bibinfo {volume} {116}},\ \bibinfo
  {pages} {147202} (\bibinfo {year} {2016})}\BibitemShut {NoStop}%
\bibitem [{\citenamefont {Tsaturyan}\ \emph {et~al.}(2017)\citenamefont
  {Tsaturyan}, \citenamefont {Barg}, \citenamefont {Polzik},\ and\
  \citenamefont {Schliesser}}]{tsaturyan2017ultracoherent}%
  \BibitemOpen
  \bibfield  {author} {\bibinfo {author} {\bibfnamefont {Yeghishe}\
  \bibnamefont {Tsaturyan}}, \bibinfo {author} {\bibfnamefont {Andreas}\
  \bibnamefont {Barg}}, \bibinfo {author} {\bibfnamefont {Eugene~S}\
  \bibnamefont {Polzik}}, \ and\ \bibinfo {author} {\bibfnamefont {Albert}\
  \bibnamefont {Schliesser}},\ }\bibfield  {title} {\enquote {\bibinfo {title}
  {Ultracoherent nanomechanical resonators via soft clamping and dissipation
  dilution},}\ }\href@noop {} {\bibfield  {journal} {\bibinfo  {journal}
  {Nature nanotechnology}\ }\textbf {\bibinfo {volume} {12}},\ \bibinfo {pages}
  {776--783} (\bibinfo {year} {2017})}\BibitemShut {NoStop}%
\bibitem [{\citenamefont {Ghadimi}\ \emph {et~al.}(2018)\citenamefont
  {Ghadimi}, \citenamefont {Fedorov}, \citenamefont {Engelsen}, \citenamefont
  {Bereyhi}, \citenamefont {Schilling}, \citenamefont {Wilson},\ and\
  \citenamefont {Kippenberg}}]{ghadimi2018elastic}%
  \BibitemOpen
  \bibfield  {author} {\bibinfo {author} {\bibfnamefont {Amir~H}\ \bibnamefont
  {Ghadimi}}, \bibinfo {author} {\bibfnamefont {Sergey~A}\ \bibnamefont
  {Fedorov}}, \bibinfo {author} {\bibfnamefont {Nils~J}\ \bibnamefont
  {Engelsen}}, \bibinfo {author} {\bibfnamefont {Mohammad~J}\ \bibnamefont
  {Bereyhi}}, \bibinfo {author} {\bibfnamefont {Ryan}\ \bibnamefont
  {Schilling}}, \bibinfo {author} {\bibfnamefont {Dalziel~J}\ \bibnamefont
  {Wilson}}, \ and\ \bibinfo {author} {\bibfnamefont {Tobias~J}\ \bibnamefont
  {Kippenberg}},\ }\bibfield  {title} {\enquote {\bibinfo {title} {Elastic
  strain engineering for ultralow mechanical dissipation},}\ }\href@noop {}
  {\bibfield  {journal} {\bibinfo  {journal} {Science}\ }\textbf {\bibinfo
  {volume} {360}},\ \bibinfo {pages} {764--768} (\bibinfo {year}
  {2018})}\BibitemShut {NoStop}%
\bibitem [{\citenamefont {Guo}\ \emph {et~al.}(2019)\citenamefont {Guo},
  \citenamefont {Norte},\ and\ \citenamefont
  {Gr{\"o}blacher}}]{guo2019feedback}%
  \BibitemOpen
  \bibfield  {author} {\bibinfo {author} {\bibfnamefont {Jingkun}\ \bibnamefont
  {Guo}}, \bibinfo {author} {\bibfnamefont {Richard}\ \bibnamefont {Norte}}, \
  and\ \bibinfo {author} {\bibfnamefont {Simon}\ \bibnamefont
  {Gr{\"o}blacher}},\ }\bibfield  {title} {\enquote {\bibinfo {title} {Feedback
  cooling of a room temperature mechanical oscillator close to its motional
  ground state},}\ }\href@noop {} {\bibfield  {journal} {\bibinfo  {journal}
  {Physical review letters}\ }\textbf {\bibinfo {volume} {123}},\ \bibinfo
  {pages} {223602} (\bibinfo {year} {2019})}\BibitemShut {NoStop}%
\bibitem [{\citenamefont {Galinskiy}\ \emph {et~al.}(2020)\citenamefont
  {Galinskiy}, \citenamefont {Tsaturyan}, \citenamefont {Parniak},\ and\
  \citenamefont {Polzik}}]{galinskiy2020phonon}%
  \BibitemOpen
  \bibfield  {author} {\bibinfo {author} {\bibfnamefont {Ivan}\ \bibnamefont
  {Galinskiy}}, \bibinfo {author} {\bibfnamefont {Yeghishe}\ \bibnamefont
  {Tsaturyan}}, \bibinfo {author} {\bibfnamefont {Micha{\l}}\ \bibnamefont
  {Parniak}}, \ and\ \bibinfo {author} {\bibfnamefont {Eugene~S}\ \bibnamefont
  {Polzik}},\ }\bibfield  {title} {\enquote {\bibinfo {title} {Phonon counting
  thermometry of an ultracoherent membrane resonator near its motional ground
  state},}\ }\href@noop {} {\bibfield  {journal} {\bibinfo  {journal} {Optica}\
  }\textbf {\bibinfo {volume} {7}},\ \bibinfo {pages} {718--725} (\bibinfo
  {year} {2020})}\BibitemShut {NoStop}%
\bibitem [{\citenamefont {Thompson}\ \emph {et~al.}(2008)\citenamefont
  {Thompson}, \citenamefont {Zwickl}, \citenamefont {Jayich}, \citenamefont
  {Marquardt}, \citenamefont {Girvin},\ and\ \citenamefont
  {Harris}}]{thompson2008strong}%
  \BibitemOpen
  \bibfield  {author} {\bibinfo {author} {\bibfnamefont {JD}~\bibnamefont
  {Thompson}}, \bibinfo {author} {\bibfnamefont {BM}~\bibnamefont {Zwickl}},
  \bibinfo {author} {\bibfnamefont {AM}~\bibnamefont {Jayich}}, \bibinfo
  {author} {\bibfnamefont {Florian}\ \bibnamefont {Marquardt}}, \bibinfo
  {author} {\bibfnamefont {SM}~\bibnamefont {Girvin}}, \ and\ \bibinfo {author}
  {\bibfnamefont {JGE}\ \bibnamefont {Harris}},\ }\bibfield  {title} {\enquote
  {\bibinfo {title} {Strong dispersive coupling of a high-finesse cavity to a
  micromechanical membrane},}\ }\href@noop {} {\bibfield  {journal} {\bibinfo
  {journal} {Nature}\ }\textbf {\bibinfo {volume} {452}},\ \bibinfo {pages}
  {72--75} (\bibinfo {year} {2008})}\BibitemShut {NoStop}%
\bibitem [{\citenamefont {Higginbotham}\ \emph {et~al.}(2018)\citenamefont
  {Higginbotham}, \citenamefont {Burns}, \citenamefont {Urmey}, \citenamefont
  {Peterson}, \citenamefont {Kampel}, \citenamefont {Brubaker}, \citenamefont
  {Smith}, \citenamefont {Lehnert},\ and\ \citenamefont
  {Regal}}]{higginbotham2018harnessing}%
  \BibitemOpen
  \bibfield  {author} {\bibinfo {author} {\bibfnamefont {Andrew~P}\
  \bibnamefont {Higginbotham}}, \bibinfo {author} {\bibfnamefont
  {PS}~\bibnamefont {Burns}}, \bibinfo {author} {\bibfnamefont
  {MD}~\bibnamefont {Urmey}}, \bibinfo {author} {\bibfnamefont
  {RW}~\bibnamefont {Peterson}}, \bibinfo {author} {\bibfnamefont
  {NS}~\bibnamefont {Kampel}}, \bibinfo {author} {\bibfnamefont
  {BM}~\bibnamefont {Brubaker}}, \bibinfo {author} {\bibfnamefont
  {G}~\bibnamefont {Smith}}, \bibinfo {author} {\bibfnamefont {KW}~\bibnamefont
  {Lehnert}}, \ and\ \bibinfo {author} {\bibfnamefont {CA}~\bibnamefont
  {Regal}},\ }\bibfield  {title} {\enquote {\bibinfo {title} {Harnessing
  electro-optic correlations in an efficient mechanical converter},}\
  }\href@noop {} {\bibfield  {journal} {\bibinfo  {journal} {Nature Physics}\
  }\textbf {\bibinfo {volume} {14}},\ \bibinfo {pages} {1038--1042} (\bibinfo
  {year} {2018})}\BibitemShut {NoStop}%
\bibitem [{\citenamefont {Stetefeld}\ \emph {et~al.}(2016)\citenamefont
  {Stetefeld}, \citenamefont {McKenna},\ and\ \citenamefont
  {Patel}}]{stetefeld2016dynamic}%
  \BibitemOpen
  \bibfield  {author} {\bibinfo {author} {\bibfnamefont {J{\"o}rg}\
  \bibnamefont {Stetefeld}}, \bibinfo {author} {\bibfnamefont {Sean~A}\
  \bibnamefont {McKenna}}, \ and\ \bibinfo {author} {\bibfnamefont {Trushar~R}\
  \bibnamefont {Patel}},\ }\bibfield  {title} {\enquote {\bibinfo {title}
  {Dynamic light scattering: a practical guide and applications in biomedical
  sciences},}\ }\href@noop {} {\bibfield  {journal} {\bibinfo  {journal}
  {Biophysical reviews}\ }\textbf {\bibinfo {volume} {8}},\ \bibinfo {pages}
  {409--427} (\bibinfo {year} {2016})}\BibitemShut {NoStop}%
\bibitem [{\citenamefont {Zhao}\ and\ \citenamefont
  {Fang}(2022)}]{zhao2022ingap}%
  \BibitemOpen
  \bibfield  {author} {\bibinfo {author} {\bibfnamefont {Mengdi}\ \bibnamefont
  {Zhao}}\ and\ \bibinfo {author} {\bibfnamefont {Kejie}\ \bibnamefont
  {Fang}},\ }\bibfield  {title} {\enquote {\bibinfo {title} {Ingap quantum
  nanophotonic integrated circuits with 1.5\% nonlinearity-to-loss ratio},}\
  }\href@noop {} {\bibfield  {journal} {\bibinfo  {journal} {Optica}\ }\textbf
  {\bibinfo {volume} {9}},\ \bibinfo {pages} {258--263} (\bibinfo {year}
  {2022})}\BibitemShut {NoStop}%
\bibitem [{\citenamefont {Verbridge}\ \emph {et~al.}(2008)\citenamefont
  {Verbridge}, \citenamefont {Ilic}, \citenamefont {Craighead},\ and\
  \citenamefont {Parpia}}]{verbridge2008size}%
  \BibitemOpen
  \bibfield  {author} {\bibinfo {author} {\bibfnamefont {Scott~S}\ \bibnamefont
  {Verbridge}}, \bibinfo {author} {\bibfnamefont {Rob}\ \bibnamefont {Ilic}},
  \bibinfo {author} {\bibfnamefont {Harold~G}\ \bibnamefont {Craighead}}, \
  and\ \bibinfo {author} {\bibfnamefont {Jeevak~M}\ \bibnamefont {Parpia}},\
  }\bibfield  {title} {\enquote {\bibinfo {title} {Size and frequency dependent
  gas damping of nanomechanical resonators},}\ }\href@noop {} {\bibfield
  {journal} {\bibinfo  {journal} {Applied Physics Letters}\ }\textbf {\bibinfo
  {volume} {93}},\ \bibinfo {pages} {013101} (\bibinfo {year}
  {2008})}\BibitemShut {NoStop}%
\bibitem [{\citenamefont {Gorodetsky}\ and\ \citenamefont
  {Grudinin}(2004)}]{gorodetsky2004fundamental}%
  \BibitemOpen
  \bibfield  {author} {\bibinfo {author} {\bibfnamefont {Michael~L}\
  \bibnamefont {Gorodetsky}}\ and\ \bibinfo {author} {\bibfnamefont {Ivan~S}\
  \bibnamefont {Grudinin}},\ }\bibfield  {title} {\enquote {\bibinfo {title}
  {Fundamental thermal fluctuations in microspheres},}\ }\href@noop {}
  {\bibfield  {journal} {\bibinfo  {journal} {JOSA B}\ }\textbf {\bibinfo
  {volume} {21}},\ \bibinfo {pages} {697--705} (\bibinfo {year}
  {2004})}\BibitemShut {NoStop}%
\bibitem [{\citenamefont {Schliesser}\ \emph {et~al.}(2008)\citenamefont
  {Schliesser}, \citenamefont {Anetsberger}, \citenamefont {Rivi{\`e}re},
  \citenamefont {Arcizet},\ and\ \citenamefont
  {Kippenberg}}]{schliesser2008high}%
  \BibitemOpen
  \bibfield  {author} {\bibinfo {author} {\bibfnamefont {Albert}\ \bibnamefont
  {Schliesser}}, \bibinfo {author} {\bibfnamefont {Georg}\ \bibnamefont
  {Anetsberger}}, \bibinfo {author} {\bibfnamefont {R{\'e}mi}\ \bibnamefont
  {Rivi{\`e}re}}, \bibinfo {author} {\bibfnamefont {Olivier}\ \bibnamefont
  {Arcizet}}, \ and\ \bibinfo {author} {\bibfnamefont {Tobias~J}\ \bibnamefont
  {Kippenberg}},\ }\bibfield  {title} {\enquote {\bibinfo {title}
  {High-sensitivity monitoring of micromechanical vibration using optical
  whispering gallery mode resonators},}\ }\href@noop {} {\bibfield  {journal}
  {\bibinfo  {journal} {New Journal of Physics}\ }\textbf {\bibinfo {volume}
  {10}},\ \bibinfo {pages} {095015} (\bibinfo {year} {2008})}\BibitemShut
  {NoStop}%
\bibitem [{\citenamefont {Walls}\ and\ \citenamefont
  {Milburn}(2007)}]{walls2007quantum}%
  \BibitemOpen
  \bibfield  {author} {\bibinfo {author} {\bibfnamefont {Daniel~F}\
  \bibnamefont {Walls}}\ and\ \bibinfo {author} {\bibfnamefont {Gerard~J}\
  \bibnamefont {Milburn}},\ }\href@noop {} {\emph {\bibinfo {title} {Quantum
  optics}}}\ (\bibinfo  {publisher} {Springer Science \& Business Media},\
  \bibinfo {year} {2007})\BibitemShut {NoStop}%
\bibitem [{\citenamefont {Uchiyama}\ \emph {et~al.}(2006)\citenamefont
  {Uchiyama}, \citenamefont {Ohta}, \citenamefont {Shiota}, \citenamefont
  {Takubo}, \citenamefont {Tanaka},\ and\ \citenamefont
  {Mochizuki}}]{uchiyama2006fabrication}%
  \BibitemOpen
  \bibfield  {author} {\bibinfo {author} {\bibfnamefont {Hiroyuki}\
  \bibnamefont {Uchiyama}}, \bibinfo {author} {\bibfnamefont {Hiroshi}\
  \bibnamefont {Ohta}}, \bibinfo {author} {\bibfnamefont {Takashi}\
  \bibnamefont {Shiota}}, \bibinfo {author} {\bibfnamefont {Chisaki}\
  \bibnamefont {Takubo}}, \bibinfo {author} {\bibfnamefont {Kenichi}\
  \bibnamefont {Tanaka}}, \ and\ \bibinfo {author} {\bibfnamefont {Kazuhiro}\
  \bibnamefont {Mochizuki}},\ }\bibfield  {title} {\enquote {\bibinfo {title}
  {Fabrication of sub-transistor via holes for small and efficient power
  amplifiers using highly selective gaas/ ingap wet etching},}\ }\href@noop {}
  {\bibfield  {journal} {\bibinfo  {journal} {Journal of Vacuum Science \&
  Technology B: Microelectronics and Nanometer Structures Processing,
  Measurement, and Phenomena}\ }\textbf {\bibinfo {volume} {24}},\ \bibinfo
  {pages} {664--668} (\bibinfo {year} {2006})}\BibitemShut {NoStop}%
\end{thebibliography}
%

\end{document}